\documentclass[12pt]{iopart}

\usepackage{iopams} 
\usepackage{graphicx} 
\usepackage{caption}  
\usepackage{subcaption} 
\usepackage{braket}

\newtheorem{theorem}{Theorem}

\begin{document}

\title[States Violating Both Locality and Noncontextuality Inequalities]{States Violating Both Locality and Noncontextuality Inequalities in Quantum Theory}

\author{Yuichiro Kitajima$^1$}
\address{$^1$ College of Industrial Technology, Nihon University

2-11-1 Shin-ei, Narashino, Chiba 275-8576, Japan}
\ead{kitajima.yuichirou@nihon-u.ac.jp} 

\begin{abstract}
The Clauser-Horne-Shimony-Holt (CHSH) inequality tests locality in quantum theory and is recognized as one of Bell's inequalities. In contrast, the Klyachko-Can-Binicio{\u{g}}lu-Shumovsky (KCBS) inequality tests noncontextuality, which refers to the idea that measurement outcomes are determined by intrinsic properties independent of other measurements that could be performed simultaneously.

While individual quantum states can violate these inequalities separately, it was believed that no single state could violate both simultaneously. This relationship suggests a trade-off between nonlocality and contextuality: if one is violated, the other cannot be. Such a relationship is known as a monogamy relation. Xue et al. demonstrated, however, that simultaneous violations of both the CHSH and KCBS inequalities are possible with specific choices of observables. This finding challenges the universal validity of the monogamy relation. They further showed that numerous scenarios exist in which both the CHSH inequality and a noncontextuality inequality involving more than five observables are violated. Nevertheless, the number of such scenarios is finite.

In this paper, we extend these findings in two significant ways. First, we show that infinitely many scenarios exist where both the CHSH inequality and a noncontextuality inequality involving an odd number of observables are violated. This demonstrates that the monogamy relation between the CHSH inequality and the noncontextuality inequality fails in infinitely many cases. Second, we identify quantum states that violate both the KCBS inequality and a locality inequality distinct from the CHSH inequality.

Our results highlight that the so-called ``monogamy relation'' between nonlocality and contextuality is not universally valid. Instead, it appears that both phenomena can coexist under broader conditions than initially expected, thus challenging the conventional viewpoint and paving the way for new theoretical and experimental research.

%While certain quantum states are known to violate these inequalities individually, it had been previously assumed that no state could violate both inequalities simultaneously. This assumption is encapsulated in the concept of the `monogamy relation.' It describes a trade-off between nonlocality and contextuality: the violation of one inequality typically excludes the possibility of violating the other. 
%These insights deepen our understanding of the complex relationship between nonlocality and contextuality, and open avenues for exploring state-dependent inequalities and their physical implications.
\end{abstract}


\submitto{\jpa} 

\maketitle

\section{Introduction}
\label{introduction-section}
Quantum theory describes phenomena that challenge classical intuition. Two key examples of such quantum behaviors are the violations of locality and noncontextuality. Locality is the principle that a measurement in one region is not influenced by measurements in spatially separated regions. Similarly, noncontextuality is the idea that measurement outcomes are determined by intrinsic properties, independent of other measurements that could be performed simultaneously. Quantum theory, however, predicts violations of locality and noncontextuality under specific conditions. It suggests that quantum objects might influence each other instantaneously regardless of distance (violating locality) or that their properties are not fixed until measured (violating noncontextuality). 

Locality has been studied using Bell inequalities, such as the Clauser-Horne-Shimony-Holt (CHSH) inequality \cite{bell1964einstein, clauser1969proposed, brunner2014bell}, which assumes that measurements in two spatially separated regions are independent. In quantum theory, various experiments have verified the violations of the CHSH inequality \cite{aspect1982experimental_test, giustina2015significant, hensen2015loophole, shalm2015strong, handsteiner2017cosmic, rauch2018cosmic}. 

In contrast, noncontextuality is examined through the Kochen-Specker theorem \cite{kochen1967problem, budroni2022kochen}. This theorem demonstrates that, under certain assumptions, no single set of pre-assigned measurement values can remain consistent across all possible measurement contexts in quantum theory. For a long time, Kochen-Specker violations remained theoretically important but experimentally elusive. That changed with the introduction of inequalities such as the Klyachko-Can-Binicio{\u{g}}lu-Shumovsky (KCBS) inequality, which made it possible to empirically test noncontextuality \cite{budroni2022kochen, klyachko2008simple}. Unlike the CHSH inequality, which involves measuring two spatially separated systems, the KCBS inequality is tested on a single quantum system by examining a particular arrangement of measurements. Experiments have shown that quantum systems do indeed violate the KCBS inequality, confirming that their measurement outcomes can depend on the measurement context \cite{ahrens2013two}.
%In contrast, noncontextuality has been examined through the Kochen-Specker theorem \cite{kochen1967problem, budroni2022kochen}, which derives contradictions under the assumption of noncontextuality for observables in quantum theory. However, the Kochen-Specker theorem was not experimentally verifiable until the development of inequalities like the Klyachko-Can-Binicio{\u{g}}lu-Shumovsky (KCBS) inequality \cite{klyachko2008simple, budroni2022kochen}. The CHSH inequality considers measurements in two spatially separated systems, while the KCBS inequality considers measurements in a single system. Similarly to the CHSH inequality, the KCBS inequality has also been experimentally confirmed to be violated in quantum theory \cite{lapkiewicz2011experimental, ahrens2013two}.

%In each case, there exist states that violate the CHSH inequality and the KCBS inequality, respectively. 
%Both nonlocality and contextuality in quantum mechanics are non-classical phenomena. However, the frameworks in which these phenomena manifest are different. Nonlocality appears in entangled states of two spatially separated systems. In contrast, contextuality is considered within a single system, not two. Therefore, it is independent of entanglement. Is there a relationship between the phenomena of nonlocality and contextuality that appear in different frameworks? 

While nonlocality typically arises in experiments involving two (or more) spatially separated and entangled particles, contextuality deals with measurements on one system. As these phenomena arise in distinct experimental setups, a natural question is whether nonlocality and contextuality are fundamentally connected or entirely independent features of quantum theory.

Kurzy{\'n}ski, Cabello, and Kaszlikowski \cite{kurzynski2014fundamental} has suggested that there might be a sort of trade-off between nonlocality and contextuality. It means that if a quantum system violates the CHSH inequality, it may be prevented from simultaneously violating the KCBS inequality, and vice versa. To show this result, they proposed two spatially separated systems. In one system, five observables, denoted as ${ B_0, B_1, B_2, B_3, B_4 }$, are measured to test the KCBS inequality. The CHSH inequality is analyzed using two observables from this system, such as $B_0$ and $B_2$, along with two from the other system.

This relationship has led to the notion of a ``monogamy relation'' between nonlocality and contextuality, suggesting that these two striking quantum features might exclude each other to some extent. In other words, a quantum state that violates the CHSH inequality might fail to violate the KCBS inequality, and vice versa, leading to a trade-off. Indeed, Kurzy{\'n}ski, Cabello, and Kaszlikowski \cite{kurzynski2014fundamental} presented a framework showing that the CHSH inequality and the KCBS inequality cannot both be violated under certain conditions, and subsequent experiments supported this conclusion \cite{zhan2016realization}.

However, Xue et al. \cite{xue2023synchronous} demonstrated that such a monogamy relation does not hold for all states. As highlighted by them, an important assumption in deriving this monogamy relation is the selection of two observables from the set ${ B_0, B_1, B_2, B_3, B_4 }$, used in the KCBS inequality, when analyzing the CHSH inequality. However, since $B_i$ and $B_{i+1}$ are simultaneously measurable observables, their product, $B_i B_{i+1}$, is also an observable. Thus, candidates for CHSH inequality observables include both $B_i$ and their products, such as $B_iB_{i+1}$.

They demonstrated that by selecting observables such as $B_0$ and $B_2 B_3$ for the CHSH inequality, there exist states that simultaneously violate both the CHSH inequality and the KCBS inequality. This finding challenges the previously accepted notion of a monogamy relation between these two inequalities. It shows that such a relation does not universally hold. The relationship between nonlocality and contextuality is more complex than this monogamy relation. Although the KCBS inequality involves five observables, they further showed that numerous scenarios exist where both the CHSH inequality and a noncontextuality inequality involving more than five observables are violated. However, the number of such scenarios is finite.

The purpose of this paper is twofold. First, it aims to demonstrate that there are infinitely many scenarios where both noncontextuality inequalities and the CHSH inequality can be violated. This finding reveals the existence of infinitely many scenarios where the monogamy relation between the CHSH inequality and the noncontextuality inequality no longer holds. Second, it seeks to show the existence of a state that violates both KCBS inequality and a locality inequality distinct from the CHSH inequality. 
%These findings open new directions for exploration and pose open problems.

Our results highlight that the so-called ``monogamy relation'' between nonlocality and contextuality is not universally valid. Instead, it appears that both phenomena can coexist under broader conditions than initially expected, thus challenging the conventional viewpoint and leading to new theoretical and experimental research.

Noncontextuality inequalities not addressed in this paper are briefly discussed here. The KCBS inequality, which tests noncontextuality in specific quantum states, is a state-dependent inequality. This means it is not violated by all states; some states do not violate this inequality. In contrast, state-independent noncontextuality inequalities are violated by all states \cite{yu2012state, huang2013experimental, budroni2022kochen}. Consequently, these inequalities are violated by any state that violates nonlocality inequalities, and the monogamy relation does not hold in a trivial sense. In this paper, such state-independent noncontextuality inequalities are not addressed.

The structure of the paper is as follows. In Section \ref{cycle-section}, a unified framework for considering the KCBS inequality and the CHSH inequality is introduced in terms of outcome independence, parameter independence, and measurement independence \cite{myrvold2024bell}. These conditions are expressed in the form of conditional probabilities concerning hidden variables, experimental settings, and experimental outcomes (Equations (\ref{OI}), (\ref{PI}), and (\ref{MI})). Outcome independence means that, under hidden variables, the outcome of one measurement does not depend on the outcome of another simultaneous measurement. Parameter independence means that under hidden variables, a measurement's outcome is unaffected by the other simultaneous measurement setting. Measurement independence means that the measurement settings are independent of the hidden variables. These conditions allow the derivation of locality inequalities (e.g., the CHSH inequality) and noncontextuality inequalities (e.g., the KCBS inequality). %In this section, the KCBS inequality is derived from slightly weaker conditions than those assumed in the original paper. 

Section \ref{CHSH-section} addresses the first objective of this paper. In this section, it is shown that for any odd number $n \geq 5$, there exist states in which both a noncontextuality inequality involving $n$ observables and the CHSH inequality are violated. Section \ref{KCBS-section} deals with the second objective. It is shown that the existence of a state that violates both the KCBS inequality and a locality inequality distinct from the CHSH inequality. This result raises an open problem. We described it in Section \ref{open-section}.

\section{CHSH inequality and KCBS inequality}
\label{cycle-section}

This section introduces the mathematical framework necessary for analyzing the CHSH and the KCBS inequalities. It provides the conditions and assumptions underlying the derivations of these inequalities.

Let $n$ be a natural number such that $n \ge 5$, let $\{ X_i \mid i = 0, \dots , n-1\}$ be a set of $n$ measurements whose values are $-1$ or $+1$, and let $x_i$ be the outcome of $X_i$.

Assume that $X_i$ and $X_{i+1}$ can be measured simultaneously, where $i+1$ and $i-1$ are taken modulo $n$. $\langle X_iX_{i+1} \rangle$ represents the expectation value when $X_i$ and $X_{i+1}$ are measured simultaneously. 
%For example, if $X_i$ and $X_{i+1}$ belong to spatially separated regions, they can be measured simultaneously.

Let $\lambda$ be a hidden variable, and impose the conditions known as outcome independence (OI), parameter independence (PI), and measurement independence (MI) on the hidden variable \cite{myrvold2024bell}. While research has been conducted to quantify these conditions in order to relax the CHSH inequality \cite{hall2010complementary, hall2011relaxed, kimura2023relaxed}, this paper does not address such quantifications.

\begin{description}
\item[Outcome Independence (OI)] 
For any $i \in \{ 0, \dots, n-1 \}$,
\begin{equation}
\label{OI}
p(x_i \vert x_{i+1}, X_i, X_{i+1}, \lambda) 
= p(x_i \vert X_i, X_{i+1}, \lambda).
\end{equation}
Outcome independence means that, under hidden variables, the outcome of one measurement does not depend on the outcome of another simultaneous measurement.

\item[Parameter Independence (PI)] 
For any $i \in \{ 0, \dots, n-1 \}$,

\begin{equation}
\label{PI}
\begin{array}{l}
p(x_i \vert X_i, X_{i+1}, \lambda) = p(x_i \vert X_i,\lambda), \\[8pt]
p(x_{i+1} \vert X_i, X_{i+1}, \lambda) = p(x_{i+1} \vert X_{i+1}, \lambda).
\end{array}
\end{equation}

Parameter independence means that under hidden variables, a measurement's outcome is unaffected by the other simultaneous measurement setting.

\item[Measurement Independence (MI)]
For any $i \in \{ 0, \dots, n-1 \}$,
\begin{equation}
\label{MI}
p(\lambda \vert X_i, X_{i+1}) = p(\lambda).
\end{equation}
Measurement independence means that the measurement settings are independent of the hidden variables.
\end{description}

When deriving the CHSH inequality, the observables considered in PI and OI were not only simultaneously observed but also spatially separated. In this section, the observables considered in PI and OI are observed simultaneously, but not necessarily spatially separated. This perspective extends the applicability of these conditions to a broader class of observables, not limited to those originally required for deriving the CHSH inequality, because observables in spatially separated regions can be treated as simultaneously measurable.

From Equations (\ref{OI}) and (\ref{PI}),
\begin{equation}
\label{product-equation}
p(x_i, x_{i+1} \vert X_i, X_{i+1}, \lambda)
= p(x_i \vert X_i,\lambda)\, p(x_{i+1} \vert X_{i+1}, \lambda).
\end{equation}

From Equations (\ref{MI}) and (\ref{product-equation}),
\begin{equation}
\label{fact}
\fl
\begin{array}{rl}
\langle X_i X_{i+1} \rangle 
&= \sum_{\lambda} \Big( p(x_i=+1, x_{i+1}=+1 \vert X_i, X_{i+1}, \lambda)\,p(\lambda \vert X_i, X_{i+1}) \\
&\quad + p(x_i=-1, x_{i+1}=-1 \vert X_i, X_{i+1}, \lambda)\,p(\lambda \vert X_i, X_{i+1}) \\
&\quad - p(x_i=+1, x_{i+1}=-1 \vert X_i, X_{i+1}, \lambda)\,p(\lambda \vert X_i, X_{i+1}) \\
&\quad - p(x_i=-1, x_{i+1}=+1 \vert X_i, X_{i+1}, \lambda)\,p(\lambda \vert X_i, X_{i+1}) \Big) \\
&= \sum_{\lambda} \Big( p(x_i=+1 \vert X_i,\lambda) 
- p(x_i=-1 \vert X_i,\lambda) \Big) \\
&\quad \times \Big( p(x_{i+1}=+1 \vert X_{i+1},\lambda)
- p(x_{i+1}=-1 \vert X_{i+1}, \lambda) \Big)\,p(\lambda).
\end{array}
\end{equation}

Using Equations (\ref{OI}) and (\ref{PI}), and (\ref{MI}), we derive Equation (\ref{fact}). This equation expresses the expectation value $\langle X_i X_{i+1} \rangle$ in terms of hidden variable probabilities and their relationships to measurement settings and outcomes.

Since

\begin{equation}
\begin{array}{rl}
\left| \sum_{i=0}^{n-2} \alpha_i \alpha_{i+1}  - \alpha_{n-1}\alpha_0 \right| \leq n-2 & (for \ even \ n) \\
 \sum_{i=0}^{n-2} \alpha_i \alpha_{i+1}  - \alpha_{n-1}\alpha_0 \leq n-2 & (for \ odd \ n)
\end{array}
\end{equation}
for any real numbers $\alpha_i$ such that $-1 \leq \alpha_i \leq 1$,

\begin{equation}
\label{inequality}
\begin{array}{rl}
\left| \sum_{i=0}^{n-2} \langle X_i X_{i+1} \rangle  - \langle X_{n-1}X_0 \rangle \right| \leq n-2 & (for \ even \ n) \\
 \sum_{i=0}^{n-2} \langle X_i X_{i+1} \rangle  - \langle X_{n-1}X_0 \rangle 
  \leq n-2 & (for \ odd \ n).
\end{array}
\end{equation}

If Inequality (\ref{inequality}) does not hold, it means that either OI, PI, or MI is violated. While some argue that MI might not hold, this paper assumes that MI does hold. From this assumption, the failure of Inequality (\ref{inequality}) implies that either OI or PI is violated. This indicates that the failure of Inequality (\ref{inequality}) is a nonlocal or contextual phenomenon.

When $n=4$, Inequality (\ref{inequality}) corresponds to the CHSH inequality, and when $n=5$, it corresponds to the KCBS inequality. The derivation of the CHSH inequality follows the standard approach, but the derivation of the KCBS inequality in this section differs from the original derivation by KCBS \cite{klyachko2008simple}. In the original KCBS derivation, it is assumed that the observable $X_i$ has a definite value of $+1$ or $-1$ whether measured simultaneously with $X_{i-1}$ or $X_{i+1}$. In other words, deterministic hidden variables are assumed.

On the other hand, in the derivation presented in this section, the value of $X_i$ is not determined even under $\lambda$, $X_i$, and $X_{i+1}$. Instead, only the probability of the value of $X_i$ is determined under $\lambda$, $X_i$, and $X_{i+1}$. That is, this section considers stochastic hidden variables. If the probability of the value of $X_i$ under $\lambda$, $X_i$, and $X_{i+1}$ is either 0 or 1, then it corresponds to deterministic hidden variables. Therefore, the KCBS inequality in this section is derived under weaker conditions compared to those assumed in the original KCBS derivation \cite{klyachko2008simple}.

According to Ara{\'u}jo et al. \cite{araujo2013all}, the CHSH inequality and the KCBS inequality can be regarded as parts of a more general framework. This framework involving $n$ observables is called an $n$-cycle scenario. In the $n$-cycle scenario, the terms where the coefficient of $\langle X_j X_{j+1} \rangle$ is $-1$ appear an odd number of times. In Inequality (\ref{inequality}), there is exactly one term where the coefficient of $\langle X_j X_{j+1} \rangle$ is $-1$, which indicates that (\ref{inequality}) represents one of the $n$-cycle scenarios. In this paper, when $n$ is even in Inequality (\ref{inequality}), we refer to it as the $n$-locality inequality. On the other hand, when $n$ is odd in Inequality (\ref{inequality}), we refer to it as the $n$-noncontextuality inequality. The $4$-locality inequality is the CHSH inequality, the $5$-noncontextuality inequality is the KCBS inequality, and both $n$-locality inequalities and $n$-noncontextuality inequalities are part of $n$-cycle scenarios.

As KCBS and Ara{\'u}jo et al. have pointed out, $n$-noncontextuality inequalities are violated in quantum theory \cite{araujo2013all}. Let $\ket{\psi_j}$ be a unit vector such that
\[ 
\ket{\psi_j}
=\frac{1}{\sqrt{1+\cos(\pi/n)}}
\left(\matrix{
 \cos j((n-1)\pi/n) \cr
 \sin  j((n-1)\pi/n) \cr 
\sqrt{\cos (\pi/n)}
}\right),
\]
%let $P_j$ be a projection such that $\ket{\psi_j}\bra{\psi_j}$,
and let
\begin{equation}
\label{kcbs-operator}
 B_j :=(-1)^{j} (2\ket{\psi_j}\bra{\psi_j} -I).
 \end{equation}
%Then  $P_jP_{j+1}=\ket{\psi_j}\braket{\psi_j | \psi_{j+1}} \bra{\psi_{j+1}}= 0$ since
 Since
 \[
 \begin{array}{rl}
\braket{\psi_j | \psi_{j+1}} 
 &=\frac{1}{1+\cos(\pi/n)}( \cos ((n-1)\pi/n) + \cos(\pi/n)) \\
 &=\frac{1}{1+\cos(\pi/n)}( - \cos (\pi/n) + \cos(\pi/n)) \\
 &=0,
 \end{array}
 \]
$\ket{\psi_j}\braket{\psi_j | \psi_{j+1}} \bra{\psi_{j+1}}= 0$. Thus, $[B_j, B_{j+1}]=0$.

 %These are observables in a spin-1 system.

Let $\ket{\varphi}$ be a unit vector such that
\begin{equation}
\label{kcbs-state}
\ket{\varphi} =
\left(
\matrix{
0 \cr
0 \cr
1
}
\right).
\end{equation}
Then
\begin{equation}
\label{kcbs-violation}
\eqalign{
\left\langle \varphi \middle| \sum_{j=0}^{n-2} B_{j}B_{j+1} - B_{2m}B_{0}  \middle| \varphi \right\rangle 
&= 4\sum_{j=0}^{n} \braket{\varphi | \psi_j}\braket{\psi_j | \varphi} -n 
\cr
&=4n\frac{\cos (\pi/n)}{1+\cos (\pi/n)} -n
\\
&>n-2.
}
\end{equation}
 
Thus, in the state represented by Equation (\ref{kcbs-state}), the $n$-noncontextuality inequality is violated for any odd $n$ such that $5 \leq n$ \cite{araujo2013all}. The observables $B_0, \dots, B_{n-1}$ will also be used in later sections.

Is there a state in which both the $n$-noncontextuality inequality and the CHSH inequality are violated for any odd $n$ such that $5 \leq n$? This question will be examined in the next section.

\section{States violating both CHSH inequality and $n$-noncontextuality inequality}
\label{CHSH-section}

In this paper, $\mathbb{C}^2$ represents a $2$-dimensional Hilbert space, while $\mathbb{C}^3$ represents a $3$-dimensional Hilbert space. In Section \ref{CHSH-section}, we consider the $n$-noncontextuality inequality in the system represented by $\mathbb{C}^3$. Using two observables employed in this inequality, we then examine the CHSH inequality in the composite system represented by $\mathbb{C}^2 \otimes \mathbb{C}^3$. 

Xue et al. \cite{xue2023synchronous} showed the existence of states that violate both KCBS inequality and CHSH inequality. Furthermore, they showed that for some odd numbers of $n$, there exist states that violate both the $n$-noncontextuality inequality and the CHSH inequality for any $n$ such that $5 \leq n \leq 10^4$. 

They analyzed specific quantum states, as described below, and demonstrated their violation of both the KCBS and CHSH inequalities. For instance, this occurs when $\theta = 0.351$, $0.421$, or $0.487$. 

\begin{equation}
\label{psi-equation}
\begin{array}{rl}
\ket{\psi} &= 
\cos \theta \left( \cos 2.868 
\left( 
\begin{array}{c}
1 \\ 
0 
\end{array} 
\right) 
+ \sin 2.868 
\left( 
\begin{array}{c}
0 \\ 
1 
\end{array} 
\right) 
\right) 
\otimes 
\left( 
\begin{array}{c}
0 \\ 
0 \\ 
1 
\end{array} 
\right) \\[8pt]
&\quad + 
\sin \theta \left( \cos 1.449 
\left( 
\begin{array}{c}
1 \\ 
0 
\end{array} 
\right) 
+ \sin 1.449 
\left( 
\begin{array}{c}
0 \\ 
1 
\end{array} 
\right) 
\right) 
\otimes 
\left( 
\begin{array}{c}
1 \\ 
0 \\ 
0 
\end{array} 
\right).
\end{array}
\end{equation}

The states that violate both inequalities exhibit two key characteristics. First, the absolute value of the coefficient $\cos \theta$ in the first term is greater than the absolute value of the coefficient $\sin \theta$ in the second term. As shown in Equation (\ref{kcbs-state}) and Inequality (\ref{kcbs-violation}), when the KCBS inequality is violated, the first term contributes significantly. Second, the state exhibits entanglement. A state that is not entangled will never violate the CHSH inequality.

Based on these two properties, Theorem \ref{CHSH-theorem} selects a state represented by a unit vector like Equation (\ref{CHSH-theorem-equation}). First, this state includes a term involving 
\[
\cos \left( \frac{\pi}{2n} \right)
\left( 
\begin{array}{c} 
0 \\ 
1 
\end{array} 
\right) 
\otimes 
\left( 
\begin{array}{c} 
0 \\ 
0 \\ 
1 
\end{array} 
\right),
\]
whose coefficient has a large absolute value. It plays a crucial role in violating the KCBS inequality. Second, this state is an entangled state.

Using the state represented by Equation (\ref{CHSH-theorem-equation}) in $\mathbb{C}^2 \otimes \mathbb{C}^3$, we show that there exist states that violate both the $n$-noncontextuality inequality for any odd $n$ such that $5 \leq n$ and the CHSH inequality in Theorem \ref{CHSH-theorem}. In this Theorem, when considering the $n$-noncontextuality inequality, the set of observables $\{ B_iB_{i+1} \mid i = 0, \dots, n-1 \}$ in $\mathbb{C}^3$ is used.
As defined in Equation (\ref{kcbs-operator}), $B_j$ is the following operator:
\[ 
\ket{\psi_j}
=\frac{1}{\sqrt{1+\cos(\pi/n)}}
\left(\matrix{
 \cos j((n-1)\pi/n) \cr
 \sin  j((n-1)\pi/n) \cr 
\sqrt{\cos (\pi/n)}
}\right),
\]
and 
\[
 B_j=(-1)^{j} (2\ket{\psi_j}\bra{\psi_j}-I).
\]

For the CHSH inequality, a subset of these observables, specifically $B_0$ and $B_m B_{m+1}$, is utilized (here, $n=2m+1$). The key point here is the choice of $B_m B_{m+1}$ rather than $B_m$ or $B_{m+1}$. For instance, in the case of $n = 5$, if $B_0$ and $B_2$ were chosen instead of $B_0$ and $B_2 B_3$, the monogamy relation \cite{kurzynski2014fundamental} would ensure that no state exists that simultaneously violates both the CHSH inequality and the KCBS inequality.

\begin{theorem}
\label{CHSH-theorem}
Let $\mathbb{C}^2$ be a two-dimensional Hilbert space and let $\mathbb{C}^3$ be a three-dimensional Hilbert space.
Let $m$ be a natural number such that $2 \leq m$, and let $n=2m+1$.

Let $\ket{\varphi}$ be a unit vector in $\mathbb{C}^2 \otimes \mathbb{C}^3$ such that
\begin{equation}
\label{CHSH-theorem-equation}
\ket{\varphi} = 
\sin \left( \frac{\pi}{2n} \right)
\left( 
\begin{array}{c} 
1 \\ 
0 
\end{array} 
\right) 
\otimes 
\left( 
\begin{array}{c} 
1 \\ 
0 \\ 
0 
\end{array} 
\right)
+ 
\cos \left( \frac{\pi}{2n} \right)
\left( 
\begin{array}{c} 
0 \\ 
1 
\end{array} 
\right) 
\otimes 
\left( 
\begin{array}{c} 
0 \\ 
0 \\ 
1 
\end{array} 
\right).
\end{equation}
Let $\{ B_j |j=0, \dots n-1 \}$ be self-adjoint operators in Equation (\ref{kcbs-operator}).
Let us define
\begin{equation}
\begin{array}{rl}
X_0 &= 
\left( 
\begin{array}{cc}
\cos \omega_0 & \sin \omega_0 \\
\sin \omega_0 & -\cos \omega_0
\end{array}
\right)
\otimes I, \\[8pt]
X_1 &= I \otimes B_{m}B_{m+1}, \\[8pt]
X_2 &= 
\left( 
\begin{array}{cc}
\cos \omega_1 & \sin \omega_1 \\
\sin \omega_1 & -\cos \omega_1
\end{array}
\right)
\otimes I, \\[8pt]
X_3 &= I \otimes B_0.
\end{array}
\end{equation}
where
\begin{equation}
\label{c0s0}
\begin{array}{rl}
c_0 &:= \frac{-2\cos(\pi/n)}{1+\cos(\pi/n)}, \\[8pt]
s_0 &:= \sin (\pi/n) \frac{2\sqrt{\cos(\pi/n)}}{1+\cos(\pi/n)}((-1)^m 2\sin(\pi/2n) - 1), \\[8pt]
c_2 &:= \frac{-4\cos(\pi/n)+2}{1+\cos(\pi/n)}, \\[8pt]
s_2 &:= \sin (\pi/n) \frac{2\sqrt{\cos(\pi/n)}}{1+\cos(\pi/n)}((-1)^m 2\sin(\pi/2n) + 1), \\[8pt]
\omega_0 &= \arctan (s_0/c_0),  \\[8pt]
\omega_2 &= \arctan (s_2/c_2).
\end{array}
\end{equation}

Then
\begin{equation}
\label{KCBS}
 \left\langle \varphi \middle| I \otimes \left( \sum_{j=0}^{n-2} B_{j}B_{j+1} - B_{n-1}B_{0} \right) \middle| \varphi \right\rangle > n-2
\end{equation}
and
\begin{equation}
\label{CHSH}
\left| \langle \varphi \middle| (X_0X_1+X_1X_2+X_2X_3-X_3X_0) \middle| \varphi \rangle \right|>2.
\end{equation}
\end{theorem}

\ref{proof-section} provides the mathematical proof of Theorem \ref{CHSH-theorem}.
By substituting a natural number greater than or equal to $2$ for $m$, the content of Theorem \ref{CHSH-theorem} becomes apparent.
 
Let $K$ be the left-hand side of Inequality (\ref{KCBS}) and let $C$ be the left-hand side of Inequality (\ref{CHSH}).
Table 1 summarizes the values of $K$ and $C$ calculated using Python for natural numbers $n$ ranging from $5$ to $13$. This table shows the violation of the $n$-noncontextuality inequality and the CHSH inequality for $n=5, 7, 9, 11, 13$.

\begin{table}[h]
\centering
\label{tab:values}
\caption{The values of $K$ and $C$ when $n$ is an odd number between $5$ and $13$. Here, $C$ represents the value of the CHSH inequality, and $K$ represents the value of the noncontextuality inequality. The value of $K$ exceeds $n-2$, and the value of $C$ exceeds $2$. Therefore, for the given $n$, both the noncontextuality inequality and the CHSH inequality are violated.}
\begin{tabular}{cccc}
\hline
\hline
 $n$ & $n - 2$ & $K$ & $C$
 \\
\hline
               5 &         3 &         3.6180 &         2.0886 \\
             7 &         5 &         5.9782 &         2.0133 \\
             9 &         7 &         8.1943 &         2.0113 \\
           11 &         9 &        10.3362 &         2.0034 \\
           13 &        11 &        12.4362 &         2.0027 \\
\hline
\hline
\end{tabular}
\end{table}

Figure 1 plots $K/n$ and $C$ on the vertical axis against $n$ on the horizontal axis for natural numbers  $n$ ranging from $5$ to $17$.

\begin{figure}[h]
    \centering
    \begin{subfigure}[t]{0.49\textwidth}
        \centering
        \includegraphics[width=\textwidth]{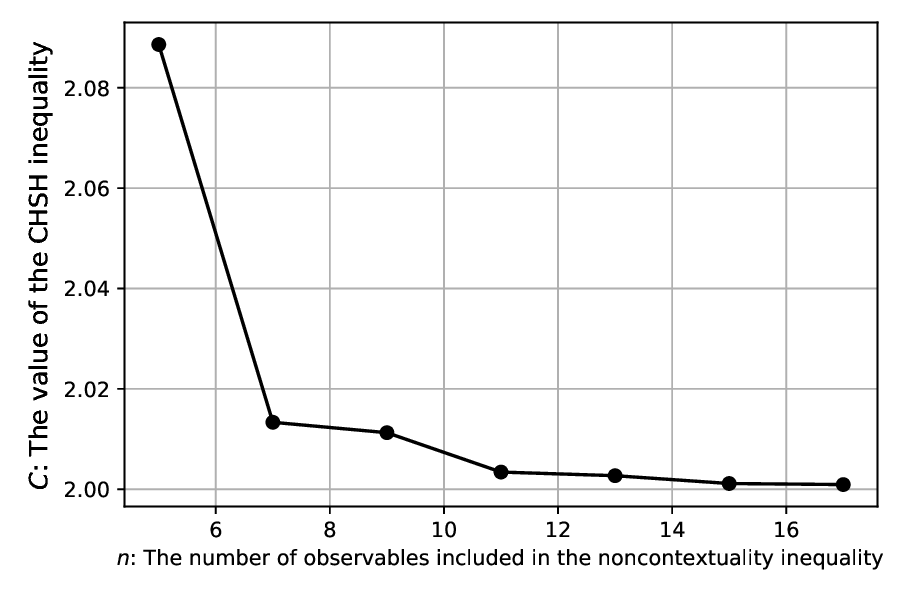}
        %\caption{CHSH Inequality}
        \label{fig:chsh}
    \end{subfigure}
    \hfill
    \begin{subfigure}[t]{0.49\textwidth}
        \centering
        \includegraphics[width=\textwidth]{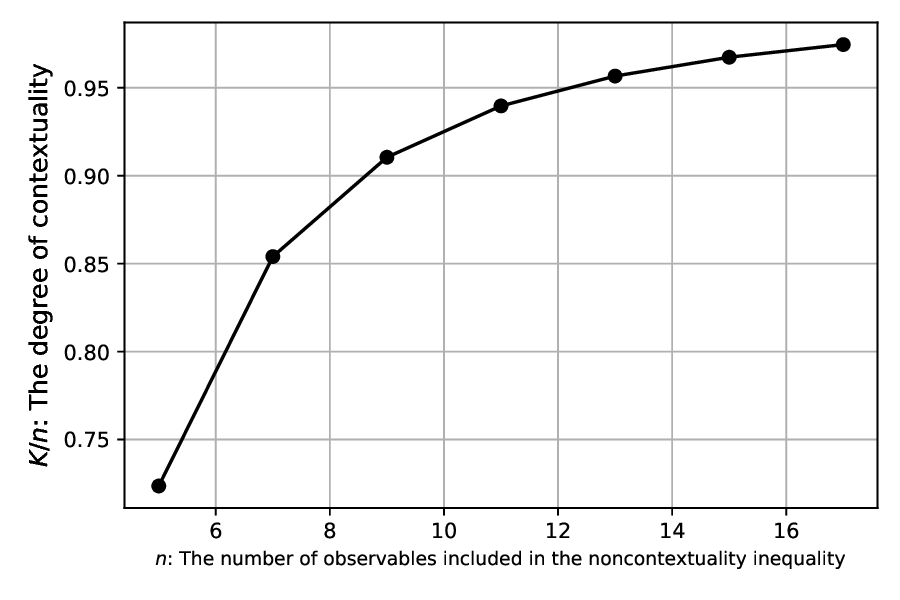}
        %\caption{Noncontextuality Inequality}
        \label{fig:noncontextuality}
    \end{subfigure}
    \caption{The left graph shows a plot with $n$ on the horizontal axis and $C$ on the vertical axis, while the right graph shows a plot with $n$ on the horizontal axis and $K/n$ on the vertical axis. Here, $C$ represents the value of the CHSH inequality, and $K$ represents the value of the noncontextuality inequality when $n$ is an odd number between $5$ and $17$. From the left graph, it can be observed that while the value of $C$ exceeds $2$, it approaches closer to $2$. From the right graph, it can be observed that the value of $K$ exceeds $n-2$, and $K$ approaches $n$. Broadly speaking, these graphs indicate that as $n$ approaches infinity, the degree of contextuality increases, while the degree of nonlocality decreases.}
    \label{fig:comparison}
\end{figure}

As shown in Equations (\ref{KCBS-equation}) and (\ref{c^2s^2}) in the proof of Theorem \ref{CHSH-theorem}, $K$ and $C$ behave differently as $n$ increases. When $n$ approaches infinity, $K$ approaches $n$. In contrast, $C$ approaches $2$. Broadly speaking, as $n$ approaches infinity, the degree of contextuality increases, while the degree of nonlocality decreases. This behavior can also be observed in Figure 1.

As stated in the proof of Theorem \ref{CHSH-theorem} in the Appendix, the value of $C$ theoretically always exceeds $2$. However, since $C$ approaches $2$ and becomes nearly identical to it, experimentally verifying whether it is greater than $2$ becomes increasingly difficult.
 
\section{States violating both $6$-locality inequality and KCBS inequality}
\label{KCBS-section}

Let $\{ B_j | j=0, \dots 4 \}$ be self-adjoint operators for $j=0, \dots 4$ defined in Equation (\ref{kcbs-operator}). When considering the CHSH inequality for the case $n = 5$ in Theorem \ref{CHSH-theorem}, we used two observables chosen from the set $\{ B_i, B_iB_{i+1} \mid i = 0, \dots, 4 \}$ within a system represented by $\mathbb{C}^3$. In this section, we explore the case where three observables are chosen instead of two. In this scenario, the CHSH inequality is not applicable.

As discussed in Section \ref{cycle-section}, there are locality inequalities involving an even number of observables besides the CHSH inequality. The $6$-locality inequality is one of them. In this section, when considering the KCBS inequality, the set of observables $\{ B_iB_{i+1} \mid i = 0, \dots, 4 \}$ in $\mathbb{C}^3$ is used. For the $6$-locality inequality, a subset of $\{ B_i, B_iB_{i+1} \mid i = 0, \dots, 4 \}$, specifically $B_0$, $B_1 B_2$ and $B_2B_3$, is utilized. Then, it is shown that there exists a state that violates both $6$-locality inequality and KCBS inequality

Let $\ket{\varphi}$ be a unit vector  in $\mathbb{C}^2 \otimes \mathbb{C}^3$ such that
\begin{equation}
\label{phi-definition}
\fl
\begin{array}{rl}
\ket{\varphi} &= 
\frac{1}{2}
\left( 
\begin{array}{c} 
1 \\ 
0 
\end{array} 
\right) 
\otimes
\left( 
\begin{array}{c} 
1 \\ 
0 \\ 
0 
\end{array} 
\right) 
+
\frac{\sqrt{3}}{2}
\left(
\frac{1}{2}
\left( 
\begin{array}{c} 
1 \\ 
0 
\end{array} 
\right) 
\otimes
\left( 
\begin{array}{c} 
0 \\ 
0 \\ 
1 
\end{array} 
\right) 
+
\frac{\sqrt{3}}{2}
\left( 
\begin{array}{c} 
0 \\ 
1 
\end{array} 
\right) 
\otimes
\left( 
\begin{array}{c} 
0 \\ 
0 \\ 
1 
\end{array} 
\right) 
\right) \\[8pt]
&= 
\frac{1}{2}
\left( 
\begin{array}{c} 
1 \\ 
0 
\end{array} 
\right) 
\otimes
\left( 
\begin{array}{c} 
1 \\ 
0 \\ 
0 
\end{array} 
\right) 
+
\frac{\sqrt{3}}{4}
\left( 
\begin{array}{c} 
1 \\ 
0 
\end{array} 
\right) 
\otimes
\left( 
\begin{array}{c} 
0 \\ 
0 \\ 
1 
\end{array} 
\right) 
+
\frac{3}{4}
\left( 
\begin{array}{c} 
0 \\ 
1 
\end{array} 
\right) 
\otimes
\left( 
\begin{array}{c} 
0 \\ 
0 \\ 
1 
\end{array} 
\right).
\end{array}
\end{equation}

The state represented by Equation (\ref{phi-definition}) is constructed similarly to the state represented by Equation (\ref{CHSH-theorem-equation}). First, this state includes a term involving
\[
\frac{\sqrt{3}}{2}
\left(
\frac{1}{2}
\left( 
\begin{array}{c} 
1 \\ 
0 
\end{array} 
\right) 
\otimes
\left( 
\begin{array}{c} 
0 \\ 
0 \\ 
1 
\end{array} 
\right) 
+
\frac{\sqrt{3}}{2}
\left( 
\begin{array}{c} 
0 \\ 
1 
\end{array} 
\right) 
\otimes
\left( 
\begin{array}{c} 
0 \\ 
0 \\ 
1 
\end{array} 
\right) 
\right)
\]
whose coefficient has a large absolute value. It strongly contributes to the violation of the KCBS inequality. Second, this state is an entangled state.

Let
\begin{equation}
\begin{array}{rl}
X_0 &= 
\left( 
\begin{array}{cc}
\cos 0 \pi & \sin 0 \pi \\
\sin 0 \pi & -\cos 0 \pi
\end{array} 
\right)
\otimes I, \\[8pt]
X_1 &= I \otimes B_1 B_2, \\[8pt]
X_2 &= 
\left( 
\begin{array}{cc}
\cos (5\pi/3) & \sin (5\pi/3) \\
\sin (5\pi/3) & -\cos (5\pi/3)
\end{array} 
\right)
\otimes I, \\[8pt]
X_3 &= I \otimes B_2 B_3, \\[8pt]
X_4 &= 
\left( 
\begin{array}{cc}
\cos (3\pi/2) & \sin (3\pi/2) \\
\sin (3\pi/2) & -\cos (3\pi/2)
\end{array} 
\right)
\otimes I, \\[8pt]
X_5 &= I \otimes B_0.
\end{array}
\end{equation}
Then
\begin{equation}
\label{6-kcbs-inequality1}
\begin{array}{rl}
&\langle \varphi | I \otimes (B_0B_1 + B_1B_2 + B_2B_3 + B_3B_4 - B_4B_0) | \varphi \rangle \\[8pt]
&= \left( \frac{1}{2} \right)^2 \left( \frac{10}{1+\cos(\pi/5)} - 5 \right) 
+ \left( \frac{\sqrt{3}}{2} \right)^2 \left( \frac{20\cos(\pi/5)}{1+\cos(\pi/5)} - 5 \right) \\[8pt]
&= \frac{5}{2} (\sqrt{5} - 1) \\[8pt]
&> 3
\end{array}
\end{equation}

and
\begin{equation}
\label{6-kcbs-inequality2}
\begin{array}{rl}
& \left| \langle \varphi \middle| X_0X_1 + X_1X_2 + X_2X_3 + X_3X_4 + X_4X_5 - X_5X_0 \middle| \varphi \rangle \right| \\[8pt]
&= \left| \frac{-36\sqrt{15} - 117\sqrt{5} + 60\sqrt{3} + 125}{80} 
+ \frac{\sqrt{\sqrt{5}+1}}{2} \left( \frac{-\sqrt{15} - 3\sqrt{5}}{5} \right) \right| \\[8pt]
&> 4.
\end{array}
\end{equation}

The values of the left-hand sides of Inequality (\ref{6-kcbs-inequality1}) and Inequality (\ref{6-kcbs-inequality2}) are approximately $3.090$ and $| -4.055 |$, respectively.

Inequalities (\ref{6-kcbs-inequality1}) and (\ref{6-kcbs-inequality2}) show that  both KCBS inequality and $6$-locality inequality are violated by the state $\ket{\varphi}$.

\section{An open problem}
\label{open-section}

In this section, we describe an open problem based on the result presented in Section \ref{KCBS-section}. In the $6$-locality inequality considered in Section \ref{KCBS-section}, three observables were considered in a system represented by $\mathbb{C}^3$. On the other hand, for the KCBS inequality, five observables were taken into account. In a system represented by $\mathbb{C}^3$, the number of observables considered for the $6$-locality inequality is fewer than the number of observables used in the KCBS inequality. When adopting the approach of using the same number of observables as in the KCBS inequality to analyze a locality inequality, it becomes necessary to employ a different inequality from the $6$-locality inequality.

For the $10$-locality inequality, instead of three observables, it is possible to select five observables from $\{ B_i, B_iB_{i+1} | i = 0, \dots, 4 \}$ in a system represented by $\mathbb{C}^3$. Under these circumstances, the problem arises of whether there exists a state that violates both the $10$-locality inequality and the KCBS inequality.

This problem can be generalized. Let $l$ be an odd number such that $l \geq 5$. The question arises whether states exist that violate both the $2l$-locality inequality and the $l$-noncontextuality inequality. For example, when $l=5$, this reduces to the previously stated problem. 

In Theorem \ref{CHSH-theorem}, when considering the KCBS inequality, a system represented in $\mathbb{C}^3$ involves $n$ observables. In contrast, when considering the CHSH inequality, only two observables from the same system represented in $\mathbb{C}^3$ are taken into account. On the other hand, in the problem mentioned above, whether analyzing the locality inequality or the noncontextuality inequality, $l$ observables are considered in a system represented in $\mathbb{C}^3$. Therefore, the problem discussed above is distinct from Theorem \ref{CHSH-theorem} and remains an open problem.

Potential resolutions to this issue include the following possibilities. For example, there might exist states that violate both the $2l$-locality inequality and the $l$-noncontextuality inequality for all $l$. Alternatively, there might exist a different type of monogamy relation, distinct from that described by Kurzy{\'n}ski, Cabello, and Kaszlikowski \cite{kurzynski2014fundamental}. In such a case, a state violating the $2l$-locality inequality might not violate the $l$-noncontextuality inequality, and vice versa. To begin with, it is necessary to examine whether a state exists that violates both the $10$-locality inequality and the KCBS inequality.

At the end of this section, we present our conjecture on this problem. We conjecture that, for sufficiently large odd $l$, there is no state that violates both the $2l$-locality inequality and the $l$-noncontextuality inequality. Theorem \ref{CHSH-theorem} provides the foundation for this conjecture.

In Figure 1 below Theorem \ref{CHSH-theorem}, consider the value $K/n$ for the $n$-noncontextuality inequality, where $K$ is its value. Similarly, for the $n$-locality inequality, consider $C/n$, where $C$ is its value. For example, in the case of $n=4$ (corresponding to the CHSH inequality), the maximum of $C/4$ is $2\sqrt{2}/4$ according to Tsirelson's bound. Under Theorem \ref{CHSH-theorem}, $K/n$ approaches $1$, while $C/n=C/4$ approaches $2/4=1/2$. This implies that as $n$ increases, the degree of contextuality grows, but the degree of locality diminishes.

Now, assume states exist that violate both the $2l$-locality inequality and the $l$-noncontextuality inequality for any odd $l$ such that $l \geq 5$. In such a case, $K/l > (l-2)/l$ and $C/2l > (2l-2)/2l$, implying that both values converge to $1$ as $l$ increases. Thus, as $l$ increases, both the degree of locality and the degree of noncontextuality also increase. This differs from Theorem \ref{CHSH-theorem}. Therefore, we conjecture that for sufficiently large odd $l$, there exists no state that violates both the $2l$-locality inequality and the $l$-noncontextuality inequality.

\section{Concluding Remarks}

As mentioned in Section \ref{introduction-section}, nonlocal phenomena arise in entangled states of two spatially separated systems, whereas contextual phenomena appear in a single system and are independent of entanglement. The significance of discovering the monogamy relation between nonlocality and contextuality was to reveal that they have a relationship although those two phenomena appear within different frameworks \cite{kurzynski2014fundamental}.

%Therefore, they seem unrelated at first glance. 
%However, Xue et al. \cite{xue2023synchronous} discovered states for which the monogamy relation does not hold. This discovery revealed that the relationship between nonlocality and contextuality is more complex than the monogamy relation suggests. Since such states were limited to a finite number, it was possible that the monogamy relation did not hold only in a finite number of special cases, and that it held in other cases. In this paper, we show that the property of simultaneously exhibiting nonlocality and contextuality is not a property that holds only in specific scenarios, but rather a generic property.

However, Xue et al. \cite{xue2023synchronous} identified specific quantum states where this monogamy relation fails. Their result indicates that the connection between nonlocality and contextuality is more intricate than previously believed. Although their analysis covered only a finite set of states, our work demonstrates that such violations occur in infinitely many scenarios. This suggests that the interplay between nonlocality and contextuality is far more complex and general than previously assumed.

In Section \ref{CHSH-section}, we showed that for any odd number $n$, there exist states that violate both the $n$-noncontextuality inequality and the CHSH inequality. This finding demonstrates the existence of infinitely many scenarios in which the monogamy relation between the CHSH inequality and the $n$-noncontextuality inequality fails.  It underscores the intricate relationship between contextuality and nonlocality, suggesting that the monogamy relation does not fully account for their interplay between them.

As described in Section \ref{cycle-section}, there are other locality inequalities involving an even number of observables besides the CHSH inequality. In Section \ref{KCBS-section}, the $6$-locality inequality involving six observables was examined, and it was shown that there exists a state violating both the $6$-locality inequality and the KCBS inequality.

These results show that the so-called ``monogamy'' between nonlocality and contextuality is not universally valid. Rather, both phenomena could coexist under a wider range of conditions than originally anticipated. It leads to new theoretical and experimental research.

The potential directions for future theoretical research were outlined in Section \ref{open-section}. The result in Section \ref{KCBS-section} raises the question of whether there exist states that violate both the $2l$-locality inequality and the $l$-noncontextuality inequality for any odd $l \geq 5$. This remains an open problem, inviting further investigation into the interplay between nonlocality and contextuality. To address this issue, it would first be necessary to investigate the case of $l=5$, that is, whether there exist states that violate both the $10$-locality inequality and the KCBS inequality.

One potential avenue for future experimental research is to verify whether the $6$-locality inequality and the KCBS inequality are simultaneously violated in the state presented in Section \ref{KCBS-section}. The simultaneous violation of the CHSH inequality and the KCBS inequality has already been experimentally demonstrated \cite{xue2023synchronous}. If the results presented in Section \ref{KCBS-section} are confirmed through this experiment, it would provide experimental evidence that the locality inequality other than the CHSH inequality can also be simultaneously violated with the KCBS inequality.

Another intriguing challenge lies in interpreting states that simultaneously violate locality and noncontextuality inequalities, particularly in terms of the underlying physical principles. As described in Section \ref{cycle-section}, these inequalities are derived from outcome independence, parameter independence, and measurement independence. The violation of both inequalities implies that at least one of these conditions fails to hold for each inequality. While there has been extensive discussion regarding the violation of the CHSH inequality \cite{myrvold2024bell}, further investigation is required for scenarios where both locality and noncontextuality inequalities are violated. Approaches that quantify outcome independence, parameter independence, and measurement independence \cite{hall2010complementary, hall2011relaxed, kimura2023relaxed} could serve as one direction for such considerations.

\appendix
\section{Proof of Theorem \ref{CHSH-theorem}}
\label{proof-section}

In this Appendix, we prove Theorem \ref{CHSH-theorem}.
\begin{enumerate}
\item First, the $n$-noncontextuality inequality is calculated. 
\begin{equation}
\label{KCBS-equation}
\begin{array}{rl}
& \left\langle \varphi \middle|  I \otimes \left( \sum_{j=0}^{n-2} B_{j}B_{j+1} - B_{n-1}B_{0} \right) \middle| \varphi \right\rangle \\[8pt]
&= \sin^2 \frac{\pi}{2n} \left( \frac{2n}{1+\cos(\pi/n)} - n \right)
+ \cos^2 \frac{\pi}{2n} \left( 4n \frac{\cos \frac{\pi}{n}}{1+\cos \frac{\pi}{n}} - n \right) \\[8pt]
&= \frac{1-\cos(\pi/n)}{1+\cos(\pi/n)} n + 2n \cos(\pi/n) - n.
\end{array}
\end{equation}
The first term in the second line of the equation uses $1+\cos^2(\pi/n)+ \dots + \cos^2((n-1)\pi/n)=n/2$, and the second term in the second line is calculated in the same manner as Equation (\ref{kcbs-violation}).

Thus
\begin{equation}
\label{kcbs-1}
\begin{array}{rl}
& \left\langle \varphi \middle| I \otimes \left( \sum_{j=0}^{n-2} B_{j}B_{j+1} - B_{n-1}B_{0} \right) \middle| \varphi \right\rangle - (n-2) \\[8pt]
&= \frac{1-\cos(\pi/n)}{1+\cos(\pi/n)}n + 2\cos(\pi/n)n - 2n + 2 \\[8pt]
&= \frac{2n\cos^2(\pi/n) + (-n+2)\cos(\pi/n) - n + 2}{1+\cos(\pi/n)}.
\end{array}
\end{equation}
Since
\begin{equation}
\label{taylor}
1-\frac{\pi^2}{2n^2} < \cos(\pi/n) < 1-\frac{\pi^2}{2n^2} +\frac{\pi^4}{24n^4}
\end{equation}
by Taylor expansion,
\begin{equation}
\label{kcbs-2}
\fl
\begin{array}{rl}
& 2n\cos^2(\pi/n) + (-n+2)\cos(\pi/n) - n + 2 \\[8pt]
&> 2n \left( 1 - \frac{\pi^2}{2n^2} \right)^2 
+ (-n+2)\left( 1 - \frac{\pi^2}{2n^2} + \frac{\pi^4}{24n^4} \right) 
- n + 2 \\[8pt]
&= 4 - \frac{3\pi^2}{2n} - \frac{\pi^2}{n^2} + \frac{11\pi^4}{24n^3} + \frac{\pi^4}{12n^4} \\[8pt]
&> 4 - \frac{3\pi^2}{2n} - \frac{\pi^2}{n^2} \\[8pt]
&> 4 - \frac{3\pi^2}{2 \times 5} - \frac{\pi^2}{5^2} \\[8pt]
&> 0
\end{array}
\end{equation}
for any natural number $n$ such that $5 \leq n$.
By Equations (\ref{kcbs-1}) and (\ref{kcbs-2}),
\[
\left\langle \varphi \middle| I \otimes \left( \sum_{j=0}^{n-2} B_{j}B_{j+1} - B_{n-1}B_{0} \right) \middle| \varphi \right\rangle
>n-2.
\]
\item Next, the CHSH inequality is calculated.

By using 
\[ \frac{2m\pi}{n}=\frac{(n-1)\pi}{n}=\pi-\frac{\pi}{n} \] and 
\[ \frac{m\pi}{n}=\frac{\pi}{2} - \frac{\pi}{2n}, \]
we obtain
\begin{equation}
\label{chsha-e1}
\fl
 \cos^2(m\pi/n)=\frac{1+\cos(2m\pi/n)}{2}=\frac{1+\cos(\pi-\pi/n)}{2}=\frac{1-\cos(\pi/n)}{2} 
\end{equation}
and
\begin{equation}
\label{chsha-e2}
\fl
\cos \left(\frac{m}{n}\pi \right)=\cos\left( \frac{1}{2} \pi - \frac{1}{2n} \pi \right)=\sin \frac{1}{2n} \pi.
\end{equation}

A detailed calculation shows that the following equality holds by Equations (\ref{chsha-e1}) and (\ref{chsha-e2}). Here, $c_0$, $c_2$, $s_0$, $s_2$, $\omega_0$, and $\omega_2$ are defined in Equation (\ref{c0s0}).
\begin{equation}
\label{CHSH-equation}
\fl
\begin{array}{rl}
& \left\langle \varphi \middle| X_0X_1 + X_1X_2 + X_2X_3 - X_3X_0 \middle| \varphi \right\rangle \\[8pt]
&= \sin^2 \frac{\pi}{2n} 
\left(
(\cos \omega_0 + \cos \omega_2) 
\left(
\frac{4\cos^2(m\pi/n)}{1+\cos(\pi/n)} - 1
\right)
+ (-\cos \omega_0 + \cos \omega_2) 
\left(
\frac{2}{1+\cos(\pi/n)} - 1
\right)
\right) \\[8pt]
&\quad + \cos^2 \frac{\pi}{2n} 
\left(
(-\cos \omega_0 - \cos \omega_2)
\left( 
\frac{4\cos(\pi/n)}{1+\cos(\pi/n)} - 1
\right) 
+ (\cos \omega_0 - \cos \omega_2)
\left( 
\frac{2\cos(\pi/n)}{1+\cos(\pi/n)} - 1
\right)
\right) \\[8pt]
&\quad + \cos \frac{\pi}{2n} \sin \frac{\pi}{2n} 
\left(
(\sin \omega_0 + \sin \omega_2)
\left( 
\frac{8}{1+\cos(\pi/n)} (-1)^m \sqrt{\cos(\pi/n)} \cos(m\pi/n)
\right)
\right) \\[8pt]
&\quad + \cos \frac{\pi}{2n} \sin \frac{\pi}{2n} 
\left(
(-\sin \omega_0 + \sin \omega_2)
\left( 
\frac{4\sqrt{\cos(\pi/n)}}{1+\cos(\pi/n)}
\right)
\right) \\[8pt]
&= \cos \omega_0
\left(
\frac{-2\cos(\pi/n)}{1+\cos(\pi/n)}
\right) \\[8pt]
&\quad + \sin \omega_0
\left(
\sin (\pi/n) \frac{2\sqrt{\cos(\pi/n)}}{1+\cos(\pi/n)} \left( (-1)^m 2 \sin(\pi/2n) - 1 \right)
\right) \\[8pt]
&\quad + \cos \omega_2
\left(
\frac{-4\cos(\pi/n) + 2}{1+\cos(\pi/n)}
\right) \\[8pt]
&\quad + \sin \omega_2
\left(
\sin (\pi/n) \frac{2\sqrt{\cos(\pi/n)}}{1+\cos(\pi/n)} \left( (-1)^m 2 \sin(\pi/2n) + 1 \right)
\right) \\[8pt]
&= c_0 \cos \omega_0 + s_0 \sin \omega_0 + c_2 \cos \omega_2 + s_2 \sin \omega_2
\end{array}
\end{equation}
 by Equation (\ref{c0s0}).

Since $\sqrt{c_i^2+s_i^2}=\sqrt{c_i^2+s_i^2}\cos(\omega_i-\omega_i)=\sqrt{c_i^2+s_i^2}\cos \omega_i \cos \omega_i + \sqrt{c_i^2+s_i^2} \sin \omega_i \sin \omega_i=c_i  \cos \omega_i+s_i \sin \omega_i$ \ ($i=0, 2$),
\begin{equation}
\label{c^2s^2}
\fl
\left\langle \varphi \middle| X_0X_1 + X_1X_2 + X_2X_3 - X_3X_0 \middle| \varphi \right\rangle
 = \sqrt{c_0^2+s_0^2}+ \sqrt{c_2^2+s_2^2} .
 \end{equation}

 Let
 \[
 \fl
\begin{array}{rl}
x_0 &:= \sqrt{\cos^2(\pi/n) + (1 - \cos (\pi/n)^2)\cos(\pi/n)\left( (-1)^m 2 \sin(\pi/2n) - 1 \right)^2}, \\[8pt]
x_2 &:= \sqrt{(-2\cos(\pi/n) + 1)^2 + (1 - \cos (\pi/n)^2)\cos(\pi/n)\left( (-1)^m 2 \sin(\pi/2n) + 1 \right)^2}, \\[8pt]
c &:= \cos(\pi/n), \\[8pt]
s &:= \sin(\pi/2n).
\end{array}
\]
Then 
\begin{equation}
\label{xcs}
\begin{array}{rl}
\sqrt{c_0^2 + s_0^2} &= 2x_0/(1+c), \\[8pt]
\sqrt{c_2^2 + s_2^2} &= 2x_2/(1+c), \\[8pt]
x_0^2 + x_2^2 &= 4c^4 - 6c^3 + c^2 + 2c + 1.
\end{array}
\end{equation}
By Inequality (\ref{c^2s^2}) and Equation (\ref{xcs}),
\[ \left\langle \varphi \middle| X_0X_1 + X_1X_2 + X_2X_3 - X_3X_0 \middle| \varphi \right\rangle  > 2 \]
is equivalent to
\[ \sqrt{c_0^2+s_0^2}+ \sqrt{c_2^2+s_2^2} > 2, \]
\[ x_0^2+x_2^2+2x_0x_2 > 1+2c+c^2, \]
\[ 4c^4-6c^3+2x_0x_2 > 0, \]
and
\[ x_0^2x_2^2 > (3c^3- 2c^4)^2. \]

Since $s^2=(1-c)/2$ and $4s=2\sqrt{2-2c}$,
\begin{equation}
\fl
\begin{array}{rl}
&x_0^2 x_2^2 - (3c^3 - 2c^4)^2 \\
&= c(c-1)^2 \big(8c^4 + 10c^3 - 3c^2 - 6c + 3  + (-1)^m 2\sqrt{2-2c} (c+1)(3c-1)\big).
\end{array}
\end{equation}

Let
\[ f(c) := 8c^4+10c^3-3c^2-6c+3 + (-1)^m 2\sqrt{2-2c} (c+1)(3c-1). \]
Then
\begin{equation}
\label{complex-inequality}
x_0^2 x_2^2 - (3c^3 - 2c^4)^2 = c(c-1)^2 f(c).
\end{equation}

Suppose that $m$ is even. Then 
\[
\begin{array}{rl}
f(c) &> 8c^4 + 10c^3 - 3c^2 - 6c + 3 \\[8pt]
&> 8\cos^4(\pi/5) + 10\cos^3(\pi/5) - 3 \times 1^2 - 6 \times 1 + 3 \\[8pt]
&= \frac{8\sqrt{5} - 7}{4} \\[8pt]
&> 0
\end{array}
\]
since $\cos(\pi/(2 \times 2 +1)) \leq c < 1$.

Suppose that $m$ is odd. Then
\[
\fl
\begin{array}{rl}
&f(c) \\[8pt]
&= 8c^4 + 10c^3 - 3c^2 - 6c + 3 - 2\sqrt{2-2c}(c+1)(3c-1) \\[8pt]
&> 8\cos^4(\pi/7) + 10\cos^3(\pi/7) - 3 \times 1^2 - 6 \times 1 + 3 - 2\sqrt{2 - 2\cos(\pi/7)}(1+1)(3 \times 1 - 1) \\[8pt]
&> 0
\end{array}
\]
since $\cos(\pi/(3 \times 2 +1)) \leq c < 1$.

Thus
\[ x_0^2x_2^2 > (3c^3- 2c^4)^2 \]
by Equation (\ref{complex-inequality}).

Therefore
\[ \left\langle \varphi \middle| X_0X_1 + X_1X_2 + X_2X_3 - X_3X_0 \middle| \varphi \right\rangle >2. \]
\end{enumerate}

\section*{Acknowledgments}
The author would like to express their gratitude to the two anonymous referees for their valuable comments.
This work was supported by JSPS KAKENHI Grant Number 20K00279.

\section*{References}
\bibliography{kitajima} 

\end{document}